\RequirePackage{fix-cm}
\documentclass[twocolumn,epjc3]{svjour3}  
\smartqed  
\RequirePackage{graphicx}
\usepackage{cite} 
\usepackage{amsmath}
\usepackage{hyperref}
\usepackage{cite}
\usepackage{amsmath,amssymb}
\usepackage{color}
\usepackage{subfigure}
\usepackage{float}
\usepackage{multicol}

 \newcommand{\dsp}{\displaystyle}
\newtheorem{thm}{Theorem}[section]

\newtheorem{ex}{Example}
\newtheorem{rem}{Remark}
\newtheorem{defn}{Definition}
\numberwithin{equation}{section}
\newcommand{\bt}{\begin{theorem}}
\newcommand{\et}{\end{theorem}}
\newcommand{\be}{\begin{equation}}
\newcommand{\ee}{\end{equation}}
\newcommand{\beqn}{\begin{eqnarray}}
\newcommand{\eeqn}{\end{eqnarray}}
\newcommand{\bqa}{\begin{eqnarray*}}
\newcommand{\eqa}{\end{eqnarray*}}

\newcommand{\bproof}{\begin{proof}}
\newcommand{\eproof}{\end{proof}}
\newcommand{\bmat}{\left ( \begin{matrix}}
\newcommand{\emat}{\end{matrix} \right )}
\newcommand{\bmatn}{\begin{matrix}}
\newcommand{\ematn}{\end{matrix} }

\begin{document}
\sloppy
\title{A vacuum solution of modified Einstein equations based on fractional calculus}


\author{A. Di Teodoro
\thanksref{e1,addr1} 
\and
E. Contreras\thanksref{e2,addr2} }
\thankstext{e1}{e-mail: 
\href{mailto:nditeodoro@usfq.edu.ec}{\nolinkurl{nditeodoro@usfq.edu.ec}}}
\thankstext{e2}{e-mail: 
\href{mailto:econtreras@usfq.edu.ec}{\nolinkurl{econtreras@usfq.edu.ec}}}

\institute{
Departamento de Matem\'atica, Colegio de Ciencias e Ingenier\'ia,\\ Universidad San Francisco de Quito USFQ, Quito 170901, Ecuador.\label{addr1}
\and
Departamento de F\'isica, Colegio de Ciencias e Ingenier\'ia,\\ Universidad San Francisco de Quito USFQ,  Quito 170901, Ecuador.\label{addr2}
}

\date{Received: date / Accepted: date}

\maketitle

\begin{abstract}
In this work, we construct a modified version of the Einstein field equations for a vacuum and spherically symmetric spacetime in terms of the Riemann-Louville fractional derivative. The main difference between our approach and other works is that we ensure that both the classical differential equations and the classical solutions are exactly recovered in the limit when the fractional parameter is turned off. We assume that the fractional equations are valid inside and near the horizon radius and match the classical solution at the horizon. Our approach resembles the Herrera--Witten strategy shown in Adv.High Energy Phys. 2018 (2018) 3839103, where the authors constructed an alternative black hole solution by assuming that inside the horizon the spacetime is hyperbolically symmetric and matches the classical spherically symmetric exterior solution at one point at the horizon. We obtain that, depending on the value of the fractional parameter, the solutions can be interpreted as a regular black hole or a gravastar. As a final step, we compute the fractional curvature scalars and show that the solution is regular everywhere inside the horizon. 
\end{abstract}

\section{Introduction}
It is a well-known fact that in general relativity (GR), several solutions of the Einstein field equations undergo curvature singularities. It is widely believed that such singularities are nonphysical and their mere presence indicates the loss of validity of the classical theory. Even more, is expected that
quantum gravitational effects become important at this scale. However, as a final quantum theory of gravitation is not consistently constructed, the interest in effective theories that account for quantum corrections has increased in recent years.

Among all the paths that can be taken to achieve an effective theory of gravity with quantum corrections, in this work, we follow the fractional calculus approach which introduces non--locality with the aim to remove the classical singularities of GR and to improve its renormalizability at the quantum level \cite{Calgani2021}. It is worth noticing that, one of the consequences of the introduction of fractional calculus to reformulate the classical GR is that spacetime becomes a non--integer dimensional entity so that the concepts of time and geometry lose their meaning but emerge only in specific regimes and approximations of the theory \cite{Calcagni2012}. This intriguing feature is encoded in different approaches 
\cite{amb2005,benedetti2009,lausher2005,horava2009,sotiriou2011} where it is found that the dimension of spacetime undergoes a flow: at high energies/small scales, the effective dimension of spacetime is two, while at lower energies the dimension runs to four, leading to classical GR \cite{Calcagni2012}. 

The change of dimensionality at different scales and its acquiring of non-integer values are typical of multi-fractals, so it is customary to describe dimensional flow as a “fractal property” of spacetime \cite{Calcagni2012}. To be more precise, all seem to indicate that the application of quantum mechanics to spacetime itself leads, in general, to a fractal geometry \cite{Calcagni2012b}. In this regard, fractional calculus is by far the most obvious tool to implement anomalous scaling in geometry.

In Ref. \cite{Calgani2021} there is presented a detailed treatment of fractional calculus in the framework of gravitation. In particular, the author proposes the classical action based on generic fractional operators and obtains the equations of motions through a variational principle. It is demonstrated that fractional Einstein equations (FEE) are formally the same as their classical counterpart, namely, they can be obtained by simply replacing the classical integer derivatives with the corresponding fractional ones. However, there is an extra ``boundary term'' whose value would depend on the choice of a certain multi--fractional measure. Nevertheless, is clear that the classical field equations are recovered in the appropriate limit (see discussion below about such a limit). It is worth mentioning that although the treatment in \cite{Calgani2021} is complete and formal, the author does not present any particular solution to the system of equations. Of course, this is not a straightforward step for two reasons: i) arriving at the Einstein field equation given a particular parameterization of the metric requires the use of a fractional Leibnitz rule which differs from the classical one by the additions of a combinatory infinite series and ii) the non--locality leads to a set of integral--differential equations which are far from trivial to solve. 

Among the first attempts to extract new physics from the FEE we can name the works by Munkhammar \cite{Munkhammar:2010gq} and Vacaru \cite{Vacaru:2010wn} in astrophysics and Roberts \cite{Roberts:2009ix} in cosmology. Of particular interest is the work \cite{Roberts:2009ix} where the author surpasses the problem of dealing with the non--trivial Leibnitz rule by imposing the so-called Last Step Modification (LSM) that consists in to replace the classical derivatives with their fractional version in the system obtained from a particular parametrization of the metric. More precisely, in \cite{Roberts:2009ix} the author replaces the time derivative appearing in the Friedmann equations with its fractional counterpart. More recently, the authors in \cite{Barrientos:2020kfp} studied a cosmological model based on the LSM and fit the fractional derivative parameters to SN Ia data to explain the current accelerated expansion of the
Universe without the use of a dark energy component. We would like to point out that the LSM must be interpreted as a route to construct effective solutions and is not a rigorous way to apply fractional calculus in general relativity. Indeed, as we commented before, applying the full machinery of fractional calculus in constructing the Einstein field equations is far from trivial and will depend on the representation of the fractional derivative we use. Alternatively, the LSM can be used as a guide to constructing the appropriate Riemann tensor by following an inverse problem strategy. In this regard, the problem reduces to finding the particular conditions that the functions have to satisfy to simplify both the Leibnitz and the chain rule of the particular fractional derivative.

In this work, we will apply the LSM technique to construct the fractional version of the Einstein equations in a vacuum, static, and, spherically symmetric spacetime whose solution corresponds to the modified Schwarzschild black hole metric by using the weighted Riemann--Liouville fractional derivative. It is worth mentioning that, our approach ensures the recovery of both the classical differential equation in terms of ordinary derivative and the classical solution which, in this case, corresponds to the Schwarzschild exterior metric. Besides, our approach resembles the Herrera--Witten strategy developed in \cite{Herrera:2018mzq}, where the authors constructed an alternative black hole solution by assuming that inside the horizon the spacetime is hyperbolically symmetric and matches the classical spherically symmetric exterior solution at one point at the horizon. The main difference here is that instead of assuming a change in the symmetry of spacetime, we consider that inside the horizon the physics is governed by fractional differential equations. 

This work is organized as follows. In the next section, we introduce the Riemann--Louville derivative. In section \ref{FEFE} we briefly introduce the Einstein field equations and construct their fractional counterpart in section \ref{FEFEQ}. In the last section, we discuss some final issues and conclude the work. In particular, we provide some comments on the use of the Riemann-Liouville operator in a broader context and discuss the shortcomings of other approaches when addressing the specific problem we tackle in this work.

\section{Riemann--Liouville}\label{RL}
This section is devoted to defining the Riemann-Liouville derivative in a rigorous way.
\\

\begin{defn} The {\bf Riemann--Liouville} fractional integral of order $\eta\in\mathbb{R}^{+}$ is given by (see \cite{KST2006, MR1993, P1999, SKM1993})
\beqn\label{IRLO}
\left(I_{b^-}^\eta h \right)(x) ~=\frac{1}{\Gamma(\eta)} \int_x^b \frac{h(t)}{(t-x)^{1-\eta}} ~dt, \quad b>x. \label{pre3}
\eeqn

\noindent We denote by $I_{b^-}^\eta(L_1)$ the class of functions $h$, represented by the fractional integral (\ref{pre3}) of a summable function, that is $h=I_{b^-}^{\eta}\varphi$, where $\varphi \in L_1(a,b).$ A description of this class of functions is given in \cite{KST2006, SKM1993}.
\end{defn}
\vspace{0.5cm}
\begin{defn}
Let $\left(D_{b^-}^\eta h \right)(x)$ denote the fractional {\bf Riemann--Liouville} derivative of order $\eta\in\mathbb{R}^{+}$ (see \cite{KST2006, MR1993, P1999,  SKM1993})
\beqn\label{fracderivative}
\left(D_{b^-}^\eta h \right)(x) &=& \nonumber \\
\left(\frac{d}{dx}\right)^s \frac{-1}{\Gamma(s-\eta)}\int_x^b \frac{h(t)}{(t-x)^{\eta-s+1}} ~dt,  \nonumber \\
s=[\eta]+1,  b>x, \label{pre1}
\eeqn
where $[\eta]$ denotes the integer part of $\eta$ and $\Gamma$ is the gamma function. When $0 <\eta <1$, then (\ref{pre1}) takes the form
\end{defn}
\beqn\label{RLO}
\left(D_{b^-}^\eta h \right)(x)~ =\frac{d}{dx} ~\frac{-1}{\Gamma(1-\eta)} \int_x^b \frac{h(t)}{(t-x)^{\eta}} ~dt. \label{pre2}
\eeqn

\begin{thm}\label{Thm1}
A function $h \in I_{b^-}^\eta(L_1), \eta>0$, if and only if, $I_{b^-}^{s-\eta} h \in AC^s([a,b])$, $s=[\eta]+1$ and
$(I_{b^-}^{s-\eta} h )^{(k)}(b)=0, \,k=0,\ldots,s-1.$ 
\label{Th1_Pre}
\end{thm}
\vspace{0.5cm}

\noindent In Theorem \ref{Thm1}, $AC^s([a,b])$ denotes the class of functions $h$, which are continuously differentiable on the segment $[a,b]$, up to order $s-1$ and $h^{(s-1)}$ is absolutely continuous on $[a,b]$. Removing the last condition in Theorem \ref{Thm1}, we get a class of functions that admits a summable fractional derivative.
\vspace{0.5cm}

\begin{rem}
We call the identity operators when $\eta=0$. That is $\left(I_{b^-}^0 h \right)(x)=h(x)$ and $\left(D_{b^-}^0 h \right)(x) = h(x)$
\end{rem}
\vspace{0.5cm}

\begin{defn}[see \cite{SKM1993}]
A function $h \in L_1(a,b)$ has a summable fractional derivative 
$\left(D_{b^-}^\eta h \right)(x)$ if
\[
\left(I_{b^-}^{s-\eta} h \right)(x) \in AC^s([a,b]),
\]
where $s=[\eta]+1.$
\end{defn}
\vspace{0.5cm}

\noindent If a function $h$ admits a summable fractional derivative,  then the composition of (\ref{pre1}) and (\ref{pre3}) can be written in the form (see \cite[Thm. 2.4]{SKM1993})
\beqn
\left( I_{b^-}^\eta ~D_{b^-}^\eta h \right)(x) = \nonumber\\
h(x) -\sum_{k=0}^{s-1} \frac{(b-x)^{\eta -k -1}}{\Gamma(\eta-k)} ~\left(I_{b^-}^{s-\eta} h \right)^{(s-k-1)} (b), \nonumber\\
 s=[\eta]+1. \label{pre8}
\eeqn

\noindent If $h \in I_{b^-}^\eta(L_1)$, then (\ref{pre8}) can be reduced to $\left( I_{b^-}^\eta ~D_{b^-}^\eta h \right)(x) =f(x)$. However, $D_{b^-}^{\eta} ~I_{b^-}^\eta h(x) = h(x)$ for both cases. This is a particular case of a more general property (see \cite[(2.114)]{P1999})
\beqn
D_{b^-}^\eta \left( I_{b^-}^\gamma h \right) = D_{b^-}^{\eta-\gamma} h, \qquad \eta \geq \gamma > 0. \label{pre8b}
\eeqn

\noindent The semigroup property for the composition of fractional derivatives does not hold in general (see \cite[Sect. 2.3.6]{P1999}). In fact, the property:
\beqn
D_{b^-}^\eta\left(D_{b^-}^\gamma h \right) =D_{b^-}^{\eta+\gamma} h \label{EL},
\eeqn
holds whenever
\beqn
h^{(j)}(b^+) =0, \qquad j=0,1, \ldots, s-1, \label{ELC}
\eeqn
and $h \in AC^{s-1}([a,b])$, $h^{(s)} \in L_1(a,b)$ and $s=[\gamma]+1$. 

In what follows, we will illustrate the implementation of the Riemann- Liouville fractional derivative with some examples.\\

\begin{ex}
If $\eta <1$ , $A\in\mathbb{R}$ and $b>x$. Then 
\begin{align}\label{ejemplo2}
D_{b^-}^\eta (x - b)^{\eta-1}=0.
\end{align}

In this regard, $(x - b)^{\eta-1}$ is a constant in the context of the Riemann-Liouville fractional derivative.
\end{ex}

\begin{ex}
If $\eta <1$ , $A\in\mathbb{R}$, $x>0$, $b>x$ 
and $h(x)=(x - b)^{1 - 2 \eta}$,
\begin{align}\label{ejemplo1}
D_{b^-}^\eta [Ah(x)]=
\frac{ A(-1)^{2-5\eta} \Gamma (2-2 \eta ) }{\Gamma (2-3 \eta )}(x-b)^{1-3 \eta },
\end{align}
\end{ex}
Note that
\begin{align} 
\lim_{\eta\to 1}D_{b^-}^\eta [Ah(x)]=\frac{3 A}{2 (x-b)^2},
\end{align}
which is proportional to the first derivative as expected. However, in this work, we insist on recovering the first derivative exactly so we propose a modification of the operator in virtue of its linearity \cite{refi,CPXNuestro20}. In this regard, we define the weighted Riemann-Liouville derivative as
\begin{eqnarray}
\mathfrak{D}_{q^-}^\eta=q(\eta)D_{b^-}^\eta,
\end{eqnarray}
with
\begin{eqnarray}
\dsp q(\eta)=\frac{A(-1)^{5\eta-1} \Gamma (2-3 \eta ) }{\Gamma (2-2 \eta )},
\end{eqnarray}
such that
\begin{eqnarray}\label{eso1}
\dsp \lim_{\eta\to 1}\mathfrak{D}_{q^-}h(x)=\frac{d}{dx}(x-b)^{-1},
\end{eqnarray}
as required. Even more, the identity map remains unchanged, namely
\begin{eqnarray}
\dsp \lim_{\eta\to 0}\mathfrak{D}_{q^-}h(x)=(x-b)\\
\end{eqnarray}
Now, by using the weighted derivative we obtain the following important results 
\begin{eqnarray}
\dsp \mathfrak{D}_{q^-}(x-b)^{\eta-1}&=&0\label{esoE1}\\
\dsp \mathfrak{D}_{q^-}(x-b)^{1-2\eta}&=&-(x-b)^{1-3\eta}\label{esoE2},
\end{eqnarray}
that we will use in future developments (see section \ref{FEFEQ}). In the next section, we will review the Einstein field equations for static and spherically symmetric spacetimes. Then, we will use the LSM method to write their fractional counterpart by using the weighted Riemann-Liouville derivative.

\section{Einstein field equations: a brief introduction}\label{FEFE}
Let us consider a spherically symmetric space-time with a line element given in Schwarzschild-like coordinates by,
\begin{eqnarray} \label{metrica}
 ds^2 = e^{\nu} dt^2 - e^{\lambda} dr^2 - r^2 \left( d\theta^{2} + \sin^{2}\theta d\phi^{2}\right),
\end{eqnarray}
where $\nu$ and $\lambda$ are functions of the radial coordinate only.
The metric (\ref{metrica}) satisfies the Einstein field equations given by,
\begin{eqnarray} \label{EFE}
 G^{\nu}_{\mu} = 8 \pi T^{\nu}_{\mu}.
\end{eqnarray}
where
\begin{eqnarray}\label{energia-momentum}
T_{\mu\nu}=(\rho+P_{\perp})u_{\mu}u_{\nu}-P_{\perp}g_{\mu\nu}+(P_{r}-P_{\perp})s_{\mu}s_{\nu},
\end{eqnarray}
encodes the matter content with, 
\begin{eqnarray}
u^{\mu}=(e^{-\nu/2},0,0,0),
\end{eqnarray}
the four-velocity of the fluid and $s^{\mu}$ is defined as
\begin{eqnarray}
s^{\mu}=(0,e^{-\lambda/2},0,0),
\end{eqnarray}
with the properties $s^{\mu}u_{\mu}=0$, $s^{\mu}s_{\mu}=-1$ (we are assuming geometric units $c=G=1$). The metric (\ref{metrica}), has to satisfy the Einstein field equations (\ref{EFE}), which are given by
\begin{eqnarray}
\rho&=&-\frac{1}{8\pi}\bigg[-\frac{1}{r^{2}}+e^{-\lambda}\left(\frac{1}{r^{2}}-\frac{\lambda'}{r}\right) \bigg],\label{ee1}\\
P_{r}&=&-\frac{1}{8\pi}\bigg[\frac{1}{r^{2}}-e^{-\lambda}\left(
\frac{1}{r^{2}}+\frac{\nu'}{r}\right)\bigg],\label{ee2}
\end{eqnarray}
\begin{equation}
P_{\perp}=\frac{1}{8\pi}\bigg[ \frac{e^{-\lambda}}{4}
\left(2\nu'' +\nu'^{2}-\lambda'\nu'+2\frac{\nu'-\lambda'}{r}
\right)\bigg]\label{ee3},
\end{equation}
where primes denote derivative with respect to $r$.

In this work, we are interested in the Schwarzschild exterior 
solution. In this case, $T_{\mu\nu}=0$ and the Einstein's equations reduce to
\begin{eqnarray}\label{eqf}
rf'+f-1=0,
\end{eqnarray}
with $f=e^{\nu}=e^{-\lambda}$. The solution of the above equation is given by
\begin{eqnarray}\label{fclasica}
f=1-\frac{2M}{r}
\end{eqnarray}
Note that this metric is `` singular'' at $r=2M$ and $r=0$. As it is well known, $r=2M$ defines the location of the event horizon of the black hole and the singularity is associated with an inadequate choice of the local chart. However, $r=0$ is a physical singularity in the sense that cannot be removed with any choice of coordinates. In this regard, sometimes is claimed that the appearance of such a singularity is a signal that the theory is not valid as $r\to0$ and that should be replaced by a suitable model. For example, given the strength of the gravitational field on and inside the horizon, it is thought that physics should be described by a quantum model of gravity. Nevertheless, as the problem of the quantization of gravity is far from being solved, some authors propose regular models supported by non--linear electrodynamics sources (not vacuum) (for an incomplete list, see \cite{Bonanno:2022rvo, Bronnikov:2022ofk,Franzin:2022wai,Konoplya:2022hll,Carballo-Rubio:2022kad,Dymnikova:2021vkb,Dymnikova:2019vuz,Dymnikova:2010zz,Bambi:2013ufa,Balart:2014cga,Ayon-Beato:1999kuh,Ayon-Beato:1998hmi}). More recently, it has been claim that, as we cannot define global static observers in the Schwarzschild background, the region $0<r<2M$ must be covered with a hyperbolic chart and, as a consequence, the singularity at $r=0$ disappear \cite{Herrera:2018mzq}. In this work, we propose that, instead of changing the symmetry inside the black hole, the singularity can be removed by writing Eintein's field equations in terms of fractional derivatives operators. It is worth mentioning that the fractional derivative introduces non--locality to the problem. Of course, this is not the first time that non--locality is introduced in order to construct a UV completed theory. Nevertheless, we will assume non--locality only inside the BH by replacing the differential operators with fractional derivatives.

\section{A simple application: Schwarzschild exterior solution}\label{FEFEQ}
Let $ds^{2}=-fdt^{2}+f^{-1}dr^{2}+r^{2}d\Omega^{2}$, be the metric of a vacuum, static, spherically symmetric and asymptotically flat space-time. Let $x=r/r_{H}$ be a dimensionless variable with $r_{H}$ the horizon radius, so Eqs. (\ref{eqf}) and (\ref{fclasica}) read
\begin{eqnarray}
x\frac{df}{dx}+f-1&=&0 \label{eqfx},\\
f&=&1-\frac{1}{x}\label{f-x}.
\end{eqnarray}
At this point, let us implement the LSM technique and define the following maps  from classical to fractional variables

\begin{eqnarray}
1&\to&(x-b)^{1-\eta}\label{map1}\\
x&\to&(x-b)^{\eta}\label{map2}\\
\frac{df}{dx}&\to&\mathfrak{D}_{q^{-}}^{\eta}\tilde{f}
\label{map3},
\end{eqnarray}
with $\tilde{f}$ the ``fractional'' function and
$b$ a dimensionless constant such that
\begin{eqnarray}
\lim\limits_{\eta\to1}b=0.
\end{eqnarray}
Using (\ref{map1})-(\ref{map3}), Eq. (\ref{eqfx}) can be written as
\begin{eqnarray}\label{eq-dif-frac}
(x-b)^{\eta}\mathfrak{D}_{q^{-}}^{\eta}\tilde{f}+\tilde{f}-(x-b)^{\eta-1}=0,
\end{eqnarray}
which corresponds to the LSM version of Eq. (\ref{eqfx}). At this point, some comments are in order. First, it should be emphasized that Eq. (\ref{eq-dif-frac}) is valid in the interval $0<x<1$ (inside the horizon, namely $0<r<2M$). Second, a solution of (\ref{eq-dif-frac}) is a vacuum solution of the fractional Einstein equations in the framework of the LSM. In this regard, spacetime is vacuum everywhere. Finally, note that  Eq. (\ref{eqfx}) is recovered when $\eta\to1$. 

\begin{figure*}
\includegraphics{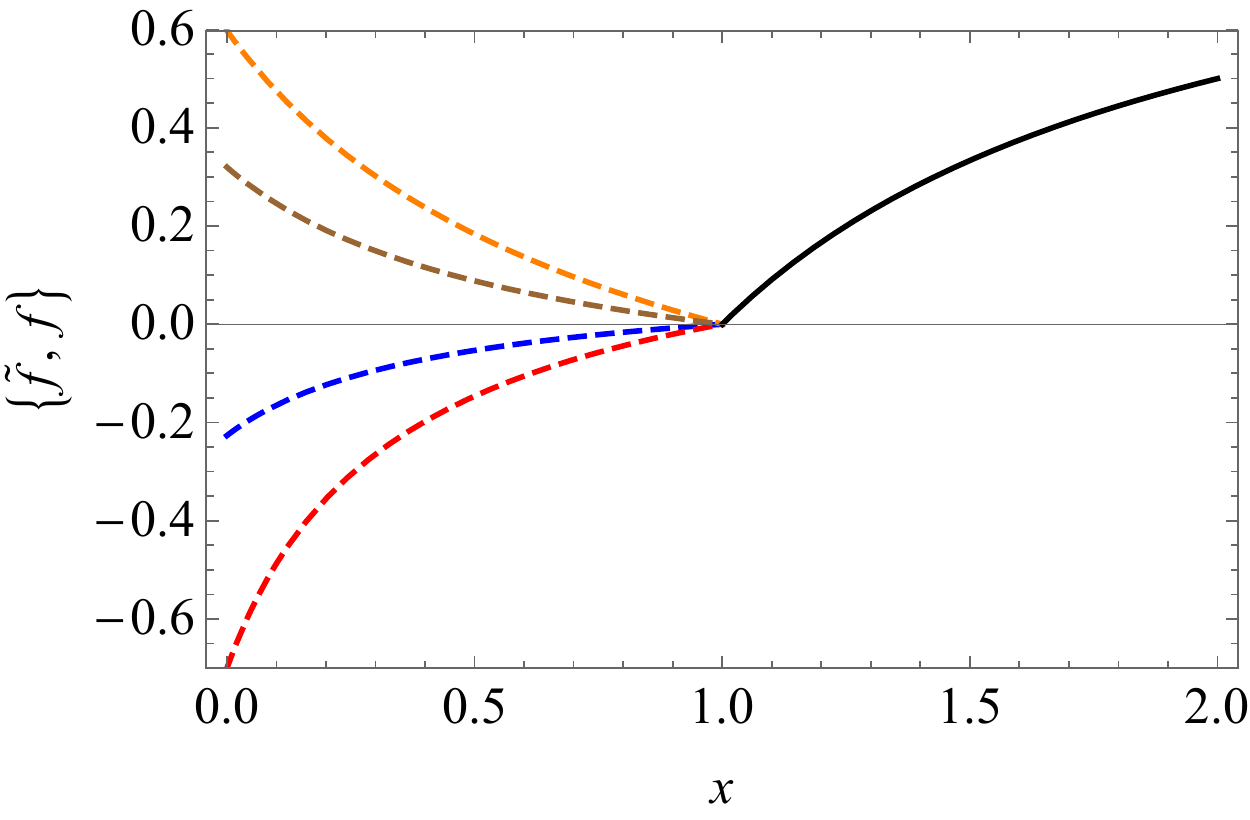}
    \caption{$\tilde{f}$ as a function of $x$ for $\alpha=0.5$ (orange line), $\alpha=0.6$ (brown line), $\alpha=0.7$ (blue line) and $\alpha=0.75$ (red line). The black line corresponds to the classical solution, $f$, which is valid for $x>1$.}
    \label{fig1}
\end{figure*}

It can be readily shown that the solution of Eq. (\ref{eq-dif-frac}) is given by
\begin{eqnarray}\label{sol-frac}
\tilde{f}=(x-b)^{\eta-1}-g(x-b)^{1-2\eta},\ \ \ g\in\mathbb{R}
\end{eqnarray}
which can be verified  by using (\ref{esoE1}) and (\ref{esoE2}). Furthermore, we are assuming  
\begin{eqnarray}
\lim\limits_{\eta\to1}g=1.
\end{eqnarray}
It is worth noticing that (\ref{sol-frac}) reduces to (\ref{f-x}), when $\eta\to1$, so the solution reduces to the classical one as expected. In order to ensure the matching of the metric at $r=r_{H}$ ($x=1$), we demand that $\tilde{f}(x=1)=0$, from where
\begin{eqnarray}
b=1-g^{\frac{1}{3\eta-2}}.
\end{eqnarray}
Although $g$ is arbitrary, in this work we shall define
\begin{eqnarray}
g=(2-\eta)^{3\eta-2},
\end{eqnarray}
from where $b=\eta-1<0$. At this point, some comments are in order. First, note that the metric is regular at $x=0$. Second, it is worth noticing that 
\begin{eqnarray}
\tilde{f}>0,\ \textnormal{if}\ \alpha<2/3 ,\\
\tilde{f}<0, \ \textnormal{if}\ \alpha>2/3 ,\\
\tilde{f}=0, \ \textnormal{if}\ \alpha=2/3.
\end{eqnarray}
In this regard, the solution mimics a ``gravastar'' \cite{mazur1,mazur2,visser,Posada:2016xxx,ovalle,raposo} for $\alpha<2/3$ and a ``regular'' black hole for $\alpha>2/3$ as shown in figure (\ref{fig1}). Interestingly, in contrast to most of the gravastars and regular black holes in literature, our model corresponds to a vacuum solution. 

\noindent It is worth mentioning that, although the metric seems regular everywhere, the computation of the scalars is compulsory, which, in accordance with our analysis, must be obtained by using the fractional derivatives. The fractional derivative version of the curvature scalars can be written after using the transformations \eqref{map1}-\eqref{map3} together with
\begin{eqnarray}
x^{-n}&\to& x^{1-(n+1)\eta},\\
x^{n}&\to&(x-b)^{n-1+\eta},\\
\frac{d^{2}f}{dx^{2}}&\to&\mathfrak{D}_{q^{-}}^{\eta}(\mathfrak{D}_{q^{-}}^{\eta}\tilde{f}),
\end{eqnarray}
with $n\ge1$. By doing so, we arrive at
\begin{eqnarray}
\tilde{R}&=&
2 x_{b}^{3 \eta -1} \left(g x_{b}^{1-2 \eta }+x_{b}^{\eta -1}-1\right)-2 g x_{b}^{1-4 \eta },\\
\widetilde{Ricc}&=&2 x_{b}^{-3} \bigg(2 g^2 x_{b}^{\eta +4}-2 g^2 x_{b}^5+x_{b}^{7 \eta },\nonumber\\
&&+x_{b}^{5 \eta +2}-2 x_{b}^{6 \eta +1}\bigg)\\
\tilde{\mathcal{K}}&=&4 x_{b}^{5 \eta -1} \bigg(g^2 x_{b}^{3-5 \eta }+g^2 x_{b}^{5-7 \eta }+\left(x_{b}^{\eta -1}-g x_{b}^{1-2 \eta }\right)^2\nonumber\\
&&-2 \left(x_{b}^{\eta -1}-g x_{b}^{1-2 \eta }\right)+1\bigg),
\end{eqnarray}
where $\tilde{R}$, $\widetilde{Ricc}$ and $\tilde{\mathcal{K}}$ stands the fractional Ricci, Ricci squared, and Kretshmann scalars respectively, and $x_{b}=x-b$. Note that the scalars are regular whenever $b>1$.

\section{Final comments and conclusion}\label{final}
In this work, we implemented fractional calculus in the framework of the Riemann-Liuville derivative to construct a static and spherically symmetric ultracompact object that mimics both regular black holes and gravastars. In contrast to what is found in the literature, our model corresponds to a vacuum solution. Indeed, the metric is the well--known Schwazschild exterior solution outside the horizon and a fractional metric inside the horizon, and both match at the event horizon radius. The interior fractional metric can be thought of as an effective solution that regularizes the singularity appearing in $r=0$ in the classical theory by taking non--local effects into account. Even more, given the strength of the gravitational field near and inside the horizon, we could conjecture that the introduction of fractional calculus accounts for quantum effects through non--locality.

It is worth mentioning that during the development of this work
we considered using other derivatives such as Caputo or Riemann Liouville on the semi-real axis, but we decided to discard them for the following reasons: 
\begin{enumerate}
    \item In the case of Caputo, the original equation of the classical case was not recovered when $\eta\to1$. Formally we could recover it under the restriction $\eta<1/2$. The main reason is that the Caputo derivative demands that the function be input differentiable, unlike the Riemann Liouville derivative. This can be seen in the definition of the fractional {Caputo} derivative of order $\eta\in\mathbb{R}, \eta>0$:

\begin{equation*}
\left(^{c}D_{a^-}^\eta h \right)(x)=
\dsp\left[\frac{1}{\Gamma (1-\eta )}\int_a^x \frac{f'(t)}{(x-t)^{\eta }} \, dt\right]
\end{equation*}
where $f'(t)=\frac{d f}{dt}$.

\item In the case of Riemann Liouville on the semi-real axis, there are no ``constants'' that can be defined. More precisely, there is not any object whose Riemann--Lioville derivative is zero.

\item 
In terms of the original non-weighted Riemann--Lioville derivative, the differential equation should read
\begin{eqnarray*}
-\frac{(x-b)^{\eta}}{(-1)^{2-5\eta}}\Big|\frac{\Gamma(2-3\eta)}{\Gamma(2-2\eta)}\Big|D_{b^{+}}^{\eta}\tilde{f}+\tilde{f}-(x-b)^{\eta-1}=0.
\end{eqnarray*}
Note that, when $\eta\to1$ the classical equation \eqref{eqf} is recovered. However, the solution of this equation is still \eqref{sol-frac} but for $\frac{\Gamma(2-3\eta)}{\Gamma(2-2\eta)}>0$, obtaining a solution for a set of reduced values of $\alpha$ and not for the entire interval.

\end{enumerate}

Before concluding this work, we would like to draw attention to the following points. First,  the Riemann-Liouville derivative presents certain challenges that could make its implementation in constructing a non-local gravity theory more difficult. For instance, constants in the regular derivative are not constants in the Riemann-Liouville derivative, which poses a problem when deriving the Einstein field equations. To be more specific, in the static and spherically symmetric case, most of the components of the Christoffel symbols vanish due to the radial dependence of both the temporal and radial components of the metric. Naturally, if we take the derivative with respect to time when constructing the Christoffel symbols, the result is zero. However, the Riemann-Liouville derivative of such a function with respect to time is non-zero, which is problematic in the context of static solutions. This issue can be addressed by imposing constraints that ensure these components become zero at the end of the computation. Another possibility could be the implementation of 
the Hilfer derivative, which is a combination of both the Caputo and Riemann-Liouville derivatives, and could potentially solve this problem.  Second, we would like to emphasize that our results here must be interpreted as an effective application of the full machinery of fractional calculus. Indeed, we could use them to apply an inverse problem strategy to constructing the Riemann tensor which leads to the fractional Einstein equation (\ref{eq-dif-frac}). Of course, it is clear that such an inverse problem will lead to very specific constraints that the function $\tilde{f}$ has to satisfy in order to simplify both the Leibnitz and the chain rule of the fractional derivative. However, we leave this and other issues to a future work.
\\

\section{Acknowledgements}
E.C is supporteb by Poligrant N$^{\circ}$ 17946.

\end{document}